\documentclass{osa-article}

\journal{oe}

\begin{document}

\title{Investigation of longitudinal spatial coherence for electromagnetic optical fields}

\author{Bhaskar Kanseri\authormark{*} and Gaytri Arya}

\address{Experimental Quantum Interferometry and Polarization (EQUIP), Department of Physics, Indian Institute of Technology Delhi, New Delhi 110016, India.}

\email{\authormark{*}bkanseri@physics.iitd.ac.in} 



\begin{abstract}
For light fields, the coherence in longitudinal direction is governed by both the frequency spectra and angular spectra they possess. In this work, we develop and report a theoretical formulation to demonstrate the effect of the angular spectra of electromagnetic light fields in quantifying their longitudinal spatial coherence. The experimental results obtained by measuring the electromagnetic longitudinal spatial coherence and degree of cross-polarization of uniformly polarized light fields for different angular spectra validate the theoretical findings.
\end{abstract}


\bibliography{Ref}

\section{Introduction}
For optical fields, the field correlation between different points in the direction of its propagation has been mainly understood as temporal coherence \cite{Wolf2007}. However, depending on the divergence of the field, coherence in the longitudinal direction can be seen as the combined effect of the frequency spectrum and angular spectrum \cite{Ryabukho2006,Ryabukho2009,Lyakin2017,abdulhalim2006,lyakin2013longitudinal,lyakin2018coherence,ryabukho2019instantaneous}. Existing studies in this direction emphasize on the role of the size of an extended source in determining longitudinal spatial coherence (LSC) \cite{Rosen1995,soroko1980,Ryabukho2004}, and use of interferometric schemes to determine LSC for scalar light fields \cite{Ryabukho2004,Lyakin2017}. It has been demonstrated that the angular diversity of the field results in amplitude-phase transformation and thus effective coherence length ($L_c$) in longitudinal direction is given in terms of temporal coherence length ($l_c$) and longitudinal spatial coherence length ($\rho_{\Vert}$) as $\frac{1}{L_C}\approx \frac{1}{l_c}+\frac{1}{\rho_{\Vert}}$\cite{Ryabukho2006,Ryabukho2009}. LSC has found several applications in applied optics \cite{soroko1980}, in radio-astronomy \cite{Ko1967} and in surface profilometry \cite{rosen2000longitudinal,duan2006dispersion}. In the medical field, it has been used to decrease effective coherence length of highly monochromatic source such as laser by orders of magnitude, providing high axial resolution in state of art imaging techniques such as optical coherence tomography (OCT) and interference microscopy \cite{abdulhalim2006, Ahmad2015,abdulhalim2012spatial,zeylikovich2008short}. Clearly, the role of LSC is quite important in controlling the coherence features of light sources.
 
 To extract the information of the tissue birefringence and scattering, and to study the electromagnetic fields, the polarization of the field also needs to be examined \cite{Tomography2009}. Since polarization and coherence both characterize the statistical similarity of correlations \cite{Wolf2007, born1999principles}, during the past decade or so, their combined effects have been explored in both spatial and temporal domains (for details see \cite{setala2006, hassinen2011, Al-qasimi2010, volkov2008, leppanen2016temporal, kanseri2010experimental,Leppanen2017, kanseri2013optical} and references therein). Since existing studies on LSC have mainly focussed on scalar aspects, and the role of vectorial nature of light has not been explored so far, it is quite essential and timely to investigate the combined effect of polarization with LSC both theoretically and experimentally. 
 
 In this paper, first we develop a theoretical framework to quantify the effect of angular spectra on the coherence in the longitudinal direction for electromagnetic (EM) fields using the standard theory of partial coherence and polarization. Our study finds that the electromagnetic degree of coherence (EMDOC) in the longitudinal direction can be written as a product of two terms: one characterizing angular spectrum and the other temporal EMDOC of the source. Further for the validation of the theory, we conduct an experiment, in which the longitudinal EMDOC has been measured for different angular spectra of a uniformly polarized source. We also determine the degree of cross-polarization (DOCP) both theoretically and experimentally for our source of variable angular spectrum.
  
\section{Theory}
Let us consider a random, statistically stationary, uniformly polarized light field propagating in z-direction emerging from a monochromatic source $\sigma$ (at $z=0$), as shown in Fig. 1. The extended source exhibits an angular spectrum of 2$\theta$ at point $P_1(0,0,z_1)$ in the subsequent plane and $\alpha$ denotes the angular coordinate due to an arbitrary point s on the source plane.
\begin{figure}[ht!]
    \centering
    \includegraphics[width=9cm,height=3cm]{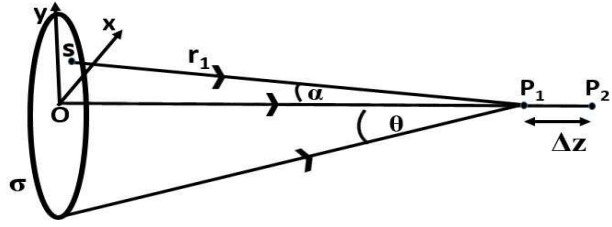}
    \caption{ Schematic diagram representing the geometrical angular spectrum (2$\theta$) of a source ($\sigma$). Symbols are described in the text.}
    \label{figg:my_label4}
\end{figure}
The electric field components reaching at point $P_m(0,0,z_m)$ (m=1, 2) from an arbitrary single point $s$ at point $P_1$ \cite{born1999principles} can be expressed as
\begin{equation}
E_i(z_m,t,\alpha)=\frac{A_i(t,\alpha)}{r_m}\exp{(\iota k r_m)};\text{for} (i=x,y),
    \end{equation}
where $'\iota'$ represents `iota' (imaginary part), $A_i(t,\alpha)$ denotes the complex amplitude at point $s$ at time $t$, $r_m$ represents the distance between points $s$ and $P_m$ (m=1, 2) and k is the wave vector. 

The coherence properties of different polarization components of light field are given by the $2\times2$ electric cross-spectral density (CSD) matrix \cite{Wolf2007}. CSD matrix $ W_{ij}(z_1,z_2,\alpha)=\langle E_i^*(z_1,t,\alpha)E_j(z_2,t,\alpha)\rangle $ describes the correlations between the field components at point $z_1$ and $z_2$ due to the point source $s$. For the extended source $\sigma$ (see sec.4.8 of \cite{soroko1980}), this matrix is obtained by integrating the angular spectra over the source plane as
\begin{equation}
     W_{ij}(z_1,z_2,\theta)=z^2 \int \int \langle E_i^*(z_1,t,\alpha)E_j(z_2,t,\alpha) d\zeta d\eta ; 
\end{equation}
where $\zeta=\frac{x}{z}$, $\eta=\frac{y}{z}$ and $\alpha=\frac{\sqrt{x^2+y^2}}{z}$ and (x,y) are the coordinates of point $s$. Using Jacobian transformation of coordinates, the integration obtained in terms of $\alpha$ is given by

\begin{equation}
    W_{ij}(z_1,z_2,\theta)=2\pi z^2\int_0^{\theta}\langle E_i^*(z_1,t,\alpha)E_j(z_2,t,\alpha)\rangle \alpha d\alpha.
\end{equation}
Considering the general case of a uniform intensity source i.e $A(t,\alpha)=A(t)$, we get the CSD matrix as 
\begin{equation}
     W_{ij}(z_m,z_n,\theta)=\frac{\pi z^2 \theta^2}{2} \exp\big({\frac{-\iota k \Delta z}{4}}\big) sinc\big(\frac{k \Delta z \theta^2}{4}\big) W_{ij}(z_m,z_n,0); (\text {for }m,n=1,2),
\end{equation}
where (m$\neq$n), and $W_{ij}(z_1,z_2,0)= \exp{(\iota k \Delta z)} \frac{\langle A_i^*(t) A_j(t)\rangle}{z^2}$ are the CSD matrix elements measuring pure temporal coherence. For $\Delta z=0$ (single point), Eq. (4) reduces to the coherence matrix. \newline
\textbf{Electromagnetic degree of coherence (EMDOC):} EMDOC quantifies coherence for EM fields \cite{setala2006}. For the case of LSC, the expression for EMDOC becomes
\begin{equation}
    \gamma_e(z_1,z_2,\theta)=\frac{tr\big[W(z_1,z_2,\theta).W(z_2,z_1,\theta)\big]}{trW(z_1,z_1,\theta)trW(z_2,z_2,\theta)}.
\end{equation}
Visibilities of generalized Stokes parameters [$V_n(z_1,z_2,\theta)$] \cite{leppanen2016temporal,setala2006} are defined as the ratio   $S_n(z_1,z_2,\theta)/S_0(z_1,z_1,\theta)$ for (n=0:3), and EMDOC can be expressed in terms of these visibilities as \cite{hassinen2011}
\begin{equation}
  \gamma_e(z_1,z_2,\theta)=\frac{1}{2}\sum_{n=0}^3\lvert V_j(z_1,z_2,\theta) \rvert ^2.
\end{equation}

On substituting CSD matrix elements from Eq. (4) into the Eq. (5), and after some rearrangement of terms, longitudinal EMDOC can be expressed as
  \begin{equation}
       \gamma_e(z_1,z_2,\theta)= \Big( \frac{2}{k \Delta z \theta^2}\Big) \sqrt{2 \Big(1-\cos{\frac{k \Delta z \theta^2}{2}}\Big)} \gamma_e(z_1,z_2,0),
  \end{equation}
where $\gamma_e(z_1,z_2,0)=(tr[W(z_1,z_2,0).W(z_2,z_1,0)])/(trW(z_1,z_1,0) trW(z_2,z_2,0))$ refers to the contribution of pure temporal EMDOC \cite{setala2006}.

Now we consider the source having Gaussian intensity distribution whose amplitude is given as $A_{i}(t,\alpha)$=$A_{i}(t)\exp{(-{\alpha^2}/{\theta^2})}$; for $i=x,y$, which is obtained on substituting r (radial distance) as $\alpha z$ and w(z)(beam waist) as $\theta z$ in Gaussian amplitude definition. Other factors of Gaussian amplitude contribute negligibly and hence can be ignored. On proceeding as above, the CSD matrix yields as
    \begin{equation}
      W_{ij}(z_1,z_2,\theta)=  \frac{\pi z^2 \theta^2} {2(1+\frac{\iota k \Delta z \theta^2}{4})}     \Big(1-\exp{\big(-2(1+\frac{\iota k \Delta z \theta^2}{4}) \big)} \Big) W_{ij}(z_1,z_2,0).
    \end{equation}
CSD matrix at point $P_1$ (m=n) can be obtain from Eq. (8) by using $\Delta z=0$. Longitudinal EMDOC obtained from this CSD matrix using Eq. (5) is given by 
\begin{equation}
    \gamma_e(z_1,z_2,\theta)= \sqrt{\frac{(1+e^{-4}-2e^{-2} \cos{\frac{k \Delta z \theta^2}{2}})}{ (1-e^{-2})^2 \big[ 1+(\frac{k \Delta z \theta^2}{4})^2\big]} } \gamma_e(z_1,z_2,0).
\end{equation}
\begin{figure}[ht!]
    \centering
    \includegraphics[width=\linewidth]{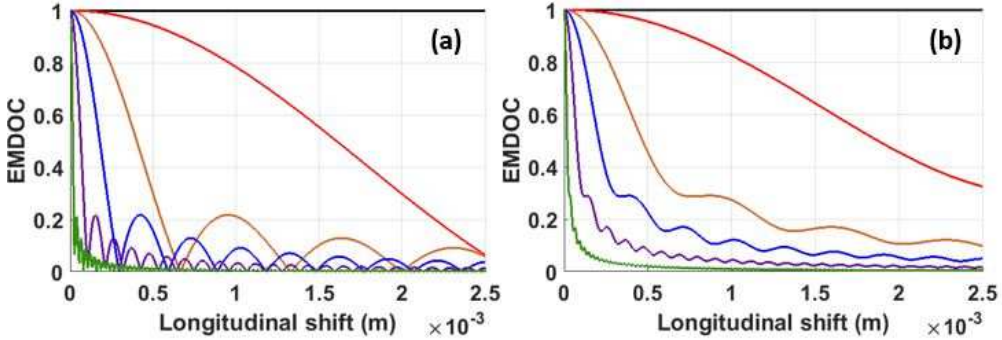}
    \caption{Plot of EMDOC vs longitudinal shift (path difference between two arms of the Michelson interferometer) for the source having (a) uniform intensity distribution [Eq. (7)], and (b) Gaussian distribution [Eq. (9)]. Colour codes for curves having angular spectrum $\theta$ (in radian): 0 (temporal) black, 0.02 red, 0.04 mustard, 0.06 blue, 0.1 purple and 0.2 green.}
    \label{figg:my_label}
\end{figure}
EMDOC is plotted for the uniform intensity distribution source and for Gaussian intensity distributed source in Fig. 2. We see that in both the cases, with increase in angular spectra, the EMDOC decreases more rapidly.  One can also notice that the dip in EMDOC for Gaussian distribution is not much pronounced compared to the case of the uniform intensity distribution. This can be due to the reason that for a Gaussian source, most of the amplitude is concentrated in the central region for Gaussian distribution (envelope effect) which produces broader maxima.

\textbf{Degree of cross polarization (DOCP):} In contrast to EMDOC [Eq. (6)], which depends on both the intensity and the polarization modulations via all the four generalized Stokes parameters, DOCP depends only on the polarization modulations of the field \cite{volkov2008}, i.e. $P(z_1,z_2,\theta)=\sqrt{\sum_{i=1}^{3}V_i(z_1,z_2,\theta)/{V_0(z_1,z_2,\theta)}}$, offering a quantification of coherence behaviour for polarization modulations only. The relation between EMDOC ($\gamma_e(z_1,z_2,\theta)$) and DOCP ($P(z_1,z_2,\theta)$) is given in \cite{hassinen2011,Al-qasimi2010,volkov2008} and for the longitudinal case it can be expressed as 
\begin{equation}
    \gamma^2_{e}(z_1,z_2,\theta)=\frac{1}{2}[1+P(z_1,z_2,\theta)]V_0(z_1,z_2,\theta).
\end{equation}
Using the values of CSD matrix from Eq. (4) in the definition of Stokes parameters \cite{setala2006} and substituting them in Eq. (10), we get DOCP (for m$\neq$n) for the field having angular spectrum (2$\theta$) equal to the DOCP for temporal field as $ P(z_m,z_n,\theta)=P(z_m,z_n,0)$. For same point $(m=n)$, DOCP reduces to degree of polarization which is identical for the diverging field and the collimated field.

\section{Experimental details}
In order to validate the theoretical findings, we determine the longitudinal EMDOC for a source having different angular spectra but uniform polarization. The experimental scheme is demonstrated in Fig. 3(a).
\begin{figure}[ht!]
    \centering
    \includegraphics[width=10cm,height=5cm]{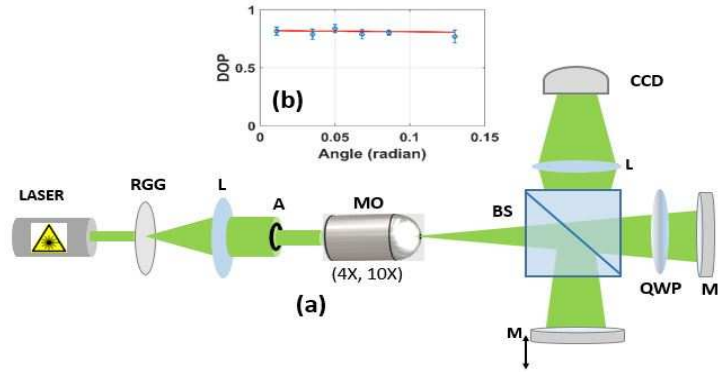}
    \caption{ (a) Scheme of the experimental set-up for investigating electromagnetic longitudinal spatial coherence. (b) Degree of polarization of the source measured for different values of the angular spectrum showing constant behaviour. Notations are described in the text.}
    \label{figg:my_label45}
\end{figure}
The spatial coherence of a vertically polarized laser beam ( $\lambda=532nm$) is destroyed by passing it through a rotating ground glass diffuser (RGG). The emerging partially coherent diverging field is collimated using a lens (L). Different angular spectra are produced with the help of an aperture (A) and microscope objectives (MOs) of numerical apertures (NAs) 4X and 10X. This field passes through a beam splitter (BS) and two mirrors (M) assembly (one mirror placed on a translation stage) and a charged coupled device (CCD) camera is used to record the interference field. A lens (L) is used just before the CCD in order to collect the diverging superimposed fields within the sensor area. The degree and state of polarization of the diffused laser beam (with and without MO) is measured using Stokes polarimetry \cite{kanseri2013optical} and is found identical for different beam divergences, as plotted and shown in Fig. 3(b). This uniformity of polarization makes the use of quarter wave plate (QWP) with the same orientation of the fast axis for different angular spectra feasible. 

The longitudinal EMDOC is determined experimentally by measuring the visibilities of four generalized Stokes parameters by converting polarization modulations into intensity modulations as reported in \cite{Leppanen2017}. The Michelson interferometric set-up shown in Fig. 3(a) is used to measure the four visibilities ($V_n$; n=0:3). No QWP is required for $V_0$ and for $V_1$, a QWP is placed in the non-translating arm with an angle of $0^0$, and for $V_2$, the QWP is rotated by $45^0$. Fig. 3(a) represents the case for the determination of $V_1$ and $V_2$. Two QWP's (one at $0^0$ and other at $45^0$) are used in the non-translating arm and one QWP at $0^o$ in translating arm are used for the measurement of $V_3$ \cite{Leppanen2017}. These visibilities are measured from the rectangular profile of the fringe pattern observed in CCD. Also, we compensate the extra path difference introduced in the non-translating arm for $V_1, V_2$ and $V_3$. The mirror placed on the translation stage is moved in a step size of 20microns and thus the path difference between the fields is shifted in the size of 40microns. 
\section{Results and discussion}
\begin{figure}[ht!]
    \centering
    \includegraphics[width=12cm,height=7cm]{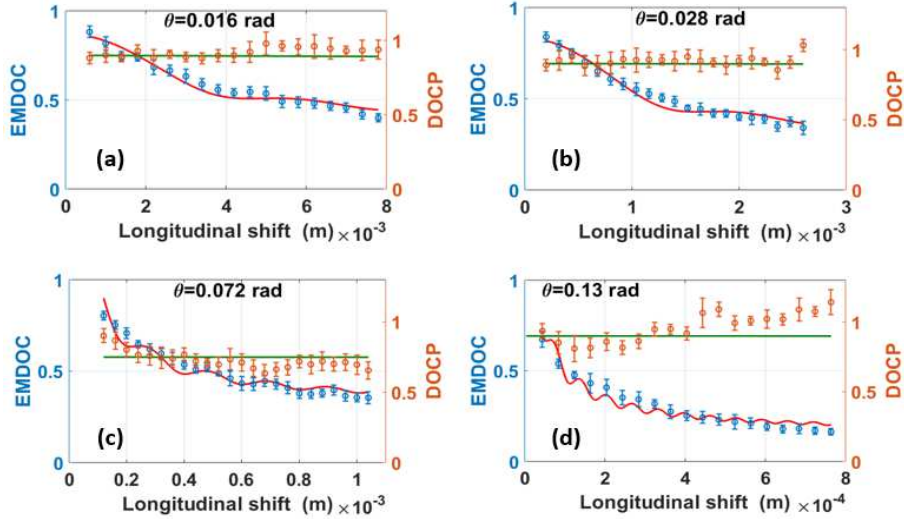}
    \caption{Plot of EMDOC (blue, left) and DOCP (orange, right) versus path difference for angular spectra ($\theta$) ranging from of 0.016rad to 0.13rad, where blue ticks are for experimental EMDOC points and red line is its fitting with Eq. (9). Orange ticks correspond to experimental DOCP points and green line represents its fitting with straight line.}
    \label{figg:my_label2}
\end{figure}
Fig. 4 shows the variation in EMDOC and DOCP with respect to the path difference in the interferometer arms, for different values of angular spectra ($\theta$ in radian) as 0.016, 0.028, 0.072 and 0.13 radian. It is apparent that the coherence length of the optical field decreases with increasing the angular spectrum whereas the frequency spectrum of the source remains unchanged. The width of the angular spectra were measured before the Michelson interferometer set-up, using a beam profiler. In order to compare these experimental results with the theoretical model proposed earlier in section 2, we fit the experimental data points of Fig. 4 with Eq.(9), using $\theta$ as a fitting parameter. The experimental uncertainty in the values of $\theta$ lies within $\pm$ 5 $\%$, demonstrating a good agreement between theory and experiment. DOCP, on the other hand, remains unchanged with the path difference. Modulations can be seen more frequent for broader angular spectra both theoretically and experimentally (Fig. 4), whereas for larger angles (0.13 radian), experimental results do not follow theoretical ones very precisely owing to the limited resolution (40 microns) of the translation stage. For the pure temporal coherence (no angular spectra, i.e. no MO), as shown in Fig. 5(a), the EMDOC varies very slowly with the path difference, as the coherence length (more than 5 meters) of the laser is decided by the frequency spectrum only. The DOCP in this case, is close to one and remains invariant with the path difference between mirrors. A comparison between Figs. 4 and 5(a) also reveals that, as derived in Eq. (9), the temporal EMDOC is constant over the path difference range of the experiment (1cm), and the effect on the coherence length is mainly due to the angular spectrum of the source.

DOCP for different angular spectra is expected to be the same as for purely temporal field, earlier explained in the Theory section. Constant behaviour of DOCP for all values of angular spectra seen in Fig. 4 is found to be similar to the constant behaviour of the temporal coherence shown in Fig. 5(a). Though the DOCP for different angular spectra behave similarly, yet the values of constants differ. DOCP can range up to infinity for correlated fields, but DOCP equals to DOP for uncorrelated fields \cite{hassinen2011}. We see that both the EMDOC and DOCP for temporally coherent field fluctuate near unity [Fig. 4(a)], whereas the values of DOCP fluctuate and drop from unity for change in angular spectra (Fig. 4). This slight discrepancy can be caused due to the experimental subtleties such as imperfect alignment and unequal intensities of interfering beams. \newline
\begin{figure}[ht!]
    \centering
    \includegraphics[width=12cm,height=4cm]{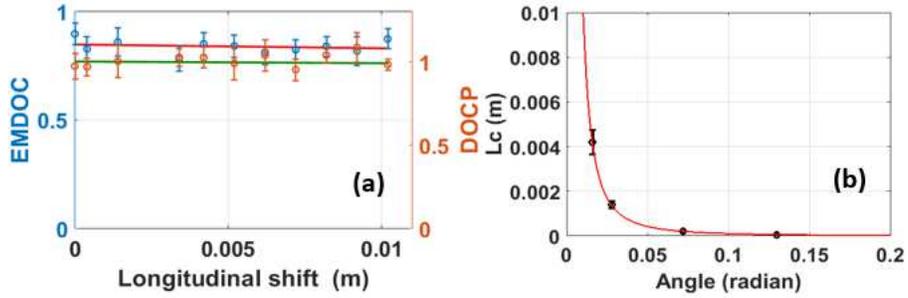}
    \caption{(a) Plot of EMDOC (left) and DOCP (right) versus path difference for purely temporal coherence (without MO), where orange ticks correspond to experimental data and green line is its straight line fit. Blue ticks denote experimentally measured EMDOC fitted using Eq. (9) with red line. (b) Plot of longitudinal coherence length ($L_c$) versus angular spectra of the source. Black ticks represent experimental data and red curve is its  theoretical fit.}
    \label{figg:my_label1}
\end{figure}
Fig. 5(b) depicts the variation in the longitudinal coherence length that can be achieved with the principle of angular diversity. The theoretical behaviour of coherence length corresponding to the given angular spectrum is plotted according to the relation $\rho_{\Vert}=(2 \lambda)/ \theta^2$ \cite{Ryabukho2009}. We clearly observe that the experimentally achieved values of coherence lengths for the angular spectra match perfectly with the theoretically expected results within measurement uncertainty. Thus one can decrease the coherence length by orders of magnitude (meters to microns), with a small increase in the angular spectra (0.016rad to 0.13rad).

\section{Conclusion}
In conclusion, we demonstrate that the coherence for diverging electromagnetic fields, in general, depends on both the temporal EM coherence and the angular dimension (divergence) of the source. Experimentally measured EM degree of coherence for different angular spectra of the field agrees well with the theoretical results. We observe that the role of polarization correlations is determined through degree of cross-polarization, which is experimentally found identical for both the diverging and non-diverging fields within the experimental uncertainties. These results are expected to be useful in characterizing the coherence properties of most general form of EM fields (such as natural light fields) which exhibit both the partial correlations and arbitrary divergence features.
   
\section{Funding}
Research leading to the results reported in this paper received funding from Science Engineering Research Board (SERB), India (grant YSS/ 2015/ 00743) and from Council of Scientific and Industrial Research (CSIR), India (grant 03(1401)/17/EMR-II). 

\end{document}